\documentstyle[12pt]{article}

\begin{document}

\title{Statistical and Scaling Properties of the $ac$ conductivity in Thin
Metal-Dielectric Composites}
\author{L. Zekri$^{1,2}$, N. Zekri$^{1,2}$, R. Bouamrane$^{1,2}$,  and
F. Brouers$^{1,3}$ \\
{\small $^{1}$International Centre for Theoretical Physics, 34100 Trieste,
Italy. }\\
{\small $^{2}$U.S.T.O., Departement de Physique, L.E.P.M., B.P.1505 El M'Naouar, 
Oran, Algeria. }\\
{\small $^{3}$Universite de Liege, Institut de Physique, Sart Tilman 4000, Liege 
Belgium.}}
\maketitle

\begin{abstract}
\hspace{0.33in} We study in this paper the scaling and statistical properties
of the $ac$ conductivity of thin metal-dielectric films in different regions of the 
loss in metallic components and particularly in
the limit of vanishing loss. We model the system by a 2D $RL-C$ network and
calculate the effective conductivity by using a real space renormalization group
method. It is found that the real conductivity strongly fluctuates 
for very small losses. The correlation length, which seems to be equivalent to the 
localization length, diverges for vanishing losses confirming our previous results 
for the decay of the real conductivity with the loss. We found also that the
distribution of the real conductivity becomes log-normal below a certain critical
loss $R_{c}$ which is size dependent for finite systems. For infinite systems
this critical loss vanishes and corresponds to the phase transition between localized
modes for finite losses and the extended ones at zero loss.
\end{abstract}

\vspace{0.2in} \noindent Keywords: Optical properties, Percolation,
Disorder, Localization, Scaling, Statistical properties.

\noindent PACS Nos. 72.15.Gd;05.70.Jk;71.55.J

\newpage
\section{Introduction}
\hspace{0.33 in} In a previous work \cite{zekri1}, we found that the real part
of the conductivity of thin metal-dielectric composites vanishes when the
loss in the metallic component vanishes. This behavior can easily be explained by
the absence of dissipation in the medium. However, this result is in
contradiction with the predictions of the effective medium theory where
\cite{Dykh}

\begin{equation}
\epsilon _{eff}=\sqrt{\epsilon _{m}\epsilon _{d}}  \label{effective}
\end{equation}
where the indices $m,d$ and $eff$ stand respectively for the metal, dielectric and
effective medium. Indeed, effective medium theory is valid only when the length
scale is larger than the correlation length \cite{Dykh,Stau} which can
eventually diverge when the loss vanishes.

\hspace{0.33 in} Recently, Brouers et al. \cite{Brou3} have studied the 
scaling effects in such films at the percolation threshold and for various
losses and frequencies. They showed the existence of two characteristic
lengths for these systems: the coherence length $L_{\omega}$ and the 
correlation length $L_{c}$. The first characteristic length corresponds to a 
ballistic transport without a modification of the phase while the second one 
corresponds to the size above which the field fluctuations become negligible.  
For small frequencies the real part of the conductivity is power-law
decreasing up to the coherence length and then increases before saturating at 
the correlation length. However, when the frequency increases the coherence
length becomes smaller than the inter-grains distance and this decrease of the
real conductivity disappears. On the other hand, they observed at the
characteristic frequency $\omega_{res}$ (where the conductivities of the two components 
have the same magnitude at vanishing loss) a mutual effect between giant
field fluctuations and the dynamic characteristic length $L_{c}$. Therefore,
the scaling behavior of the conductivity allows us to measure qualitatively
the correlation length particularly for vanishing loss where there is still a
controversy as discussed above \cite{zekri1}. We note that the scaling
study of Brouers et al. \cite{Brou3} has been restricted only for two values
of the loss: $R= 0.1$ and $R= 10^{-4}$. However, the above mentioned discrepancy
with the effective medium theory has been observed for smaller losses 
\cite{zekri1}. It is then necessary to measure $L_{c}$ for smaller losses
in order to check wether it is divergent or not in the limit
superconductor-dielectric.

\hspace{0.33 in} On the other hand, a non monotonic behavior of the real
conductivity has been observed in the previous work \cite{zekri1}  in the
region around $R=10^{-6}$ most probably due to strong fluctuations of
conductivity. This quantity, averaged over 100 samples could not behave
'{\it well}' statistically. Therefore, the competition between the scaling effect
and the statistical properties can play an important role in such systems and
in particular, the transition from localized modes below the correlation length
to extended ones above this length \cite{zekri1} can be statistically
characterized as for quantum systems. Indeed, in the quantum counter part the
conductance fluctuations change from the insulating phase to the conducting one.
In particular, in the metallic phase the conductance fluctuations are universal
and of the order of $e^2/h$ with a Gaussian distribution, while in the
insulating phase these fluctuations become large and the distribution is
log-normal \cite{Zano}.

\hspace{0.33 in} In this paper, we examine the scaling properties of this system for 
a wide range of losses including the limit superconductor-insulator and determine the 
behavior of the correlation length in this limit. We study also and characterize the 
different eigenmode phases in these systems by the statistical properties of
the conductivity for various losses. We restrict ourselves to the concentration
of the metal corresponding to the percolation threshold (which is
the minimum concentration corresponding to the appearance of a continuous
metallic path through the sample. In 2D $p_{c}=0.5$). As in the previous work 
\cite{zekri1}, this system 
is modelled by a two dimensional $RL-C$ network where the inductance $L$ stands for 
the metallic grains with a small loss (resistance) $R$ while the dielectric grains 
correspond to the capacitance $C$ which is assumed without dissipation. The major part
of the calculations is done for the characteristic frequency $\omega_{res}$ 
(At this frequency the dielectric $ac$ conductivity has the same magnitude 
in modulus as the metallic one for small losses). We can use, without loss of
generality the framework where $L$=$C$=$\omega_{res}$=$1$. Frequencies different
from the characteristic one are normalized by $\omega_{res}$. We
use for the calculation of the effective conductivity the Real Space
Renormalization Group Method (RSRG) extensively studied during the two last
decades \cite{zekri1,Brou3,Brou2,Sary1} which consists in a representation of
the network in Weatstone bridges transformed into two equivalent conductivities
following the directions $x$ and $y$ (see Fig.1). This method has been shown
be a good approximation for the calculation of the conductivity and the critical
exponents near the percolation threshold \cite{zekri1,Brou3}.

\section{Scaling properties and characteristic lengths}

\hspace {0.33 in} Since, from the previous work \cite{zekri1}, the composite
metal-dielectric films showed localization properties, then it should exist in 
addition to the two characteristic lengths found previously ($L_{\omega}$ and $L_{c}$) 
\cite{Brou3}, a third one which is the localization length $L_{loc}$. However, this 
length seems to be equivalent to $L_{c}$. Indeed, the localization length is defined
as the mean size of the sample above which the local field strength (the equivalent to 
the wavefunction in Helmoltz equation) decays exponentially and becomes negligible.
This means also that above this length the field fluctuations become small 
which is the definition of the correlation length $L_{c}$.

\hspace{0.33 in} On the other hand, we found previously \cite{zekri1}, by means
the inverse participation ratio (IPR) that the eigenmodes are delocalized when
the loss in the metallic grains tends to vanish without becoming purely extended
(the exponent of the IPR does not reach the value $-2$). This probably means
that the localization length increases but remains finite. We should then
confirm the divergence of the correlation length for vanishing losses before
discussing its relation to the localization length. We re-examine now the
scaling properties of the real part of the effective conductivity as done by
Brouers et al. \cite{Brou3} but for a range of losses extended to vanishing ones
and for two frequencies: the characteristic one($\omega_{res} = 1$) and
$\omega = 1/8$ (obviously we find the same results as Brouers et al. for the
values of the loss studied by them, i.e. $R=0.1$ and $R=10^{-4}$). As they found,  
the real conductivity increases and saturates at the correlation length 
(see Figs.2). Furthermore, for small losses the effective conductivity is shown to 
fluctuate strongly in opposition to the case of large losses where it seems to behave 
'{\it well}' statistically. We remark also (as expected) that the saturation takes 
place at larger sizes when the loss is small, and becomes much greater than $1024$ 
(which is the maximum size we can reach with our computers) below $R=10^{-6}$. 

\hspace{0.33 in} In Fig.3 we see that the correlation length increases rapidly 
when the loss decreases and is power-law diverging for vanishing losses (see the 
insert of Fig.3) with the power law exponent $0.41$ (close to $0.5$ expected by 
Brouers et al. \cite{Brou3}). This correlation length is estimated from the 
saturation of the 
effective conductivity shown in Fig.2 and is not accurate because of the strong 
fluctuations of the conductivity particularly for small losses. But this divergence 
is clearly shown from the general behavior of the effective conductivity. Therefore, 
from the above discussion, this means that the localization length diverges also. 
Indeed, although the exponent of the IPR saturates at a value smaller than $-2$, this
quantity itself (the IPR) does not saturate and continues decreasing with the
length scale. Therefore, the localization and the correlation length have at
least similar behaviors and confirm the decrease of the conductivity for
vanishing losses. 

\section{statistical behavior of the conductivity} 

\hspace{0.33 in} As we can see in Fig.2, when the loss becomes small the conductivity 
(which is averaged over 100 samples) becomes strongly fluctuating. Therefore, as 
discussed before, the conductivity could not obey to the central limit theorem 
\cite{Vent}, and we should study its statistics before averaging it.
On the other hand, for quantum systems, in the insulating phase, the conductance 
distribution is log-normal and then, its logarithm obeys to the central limit 
theorem \cite{Vent}. The conductance fluctuations for classical systems have been 
studied for a long time \cite{Esso} but have not characterized the electromagnetic 
eigenmodes. 

\hspace{0.33 in} In Figs.4, we show the distribution of the conductivity for
various losses. As we can see in these figures, in the region of large losses 
($R=10^{-1}$ and $10^{-3}$) the distribution of the real conductivity is Gaussian
and becomes narrower when the length scale increases indicating that the real
conductivity obeys to the central limit theorem \cite{Vent}. We see also that when
the loss decreases these distributions become broaden, and consequently the
conductivity fluctuations increase (as clearly shown in Fig.5). In the region
of very small losses ($R=10^{-6}$ and $10^{-9}$ in Figs.4c and 4d respectively) 
this distribution becomes Poissonian and narrows for larger sample sizes for 
$R=10^{-6}$ (Fig.4c) while it seems to be less affected by the size for $R=10^{-9}$ 
(Fig.4d). These two different behaviors of the conductivity distribution appear 
clearly in Fig.5 for the variance which decreases for $R$ smaller than $10^{-8}$ 
while the relative variance (called also normalized noise \cite{Esso}) saturates 
in this region. Therefore, the averaged real conductivity seems to be not 
'{\it stable}' statistically for very small losses, while it seems to be so in 
the region around $R=10^{-6}$.

\hspace{0.33 in} In Fig.6 we show the distribution of $-Log(\sigma)$ in the 
region of the loss where $\sigma$ shows a Poissonian distribution in Figs.4
(i.e. $R=10^{-6}$ and $10^{-9}$). We can see that the distribution for
$R=10^{-6}$ seems to become Gaussian and becomes narrower for larger system sizes
while for $R=10^{-9}$ the distribution has a long tail for large $\sigma$, but
the narrowing seems to be slowly varying with the size. From these distributions we
conclude that in this region of very small losses $Log(\sigma)$ is the relevant
quantity for statistical averaging and not $\sigma$. Note that these
distributions have been done only from $100$ samples which is probably not
sufficient for obtaining the Gaussian shape, but their narrowing for larger
sizes it is clearly observed. We estimate the critical loss separating the two
regions (Gaussian distribution and log-normal one) at $R_{c}=10^{-5}$. Therefore, 
below this critical loss the real conductivity should be estimated from the
average of its logarithm, i.e.

\begin{equation}
< \sigma > = exp \left ( <Log(\sigma)> \right )
\end{equation}

\hspace{0.33 in} In Fig.7 we compare the real conductivity calculated by direct 
averaging ($<\sigma>$) with the averaging of this quantity by Eq.(2) in the regions 
of the log-normal distributions. We see that the second method gives average effective 
conductivities for very small losses one magnitude smaller than in the first one. 
This implies that the correlation length should increase for vanishing losses more
rapidly than in Fig.3. 

\hspace{0.33 in} Finally this statistical behavior of the real conductivity
characterizes the two phases: localized modes for sizes above the correlation
length and extended modes below this length (which diverges at zero loss). This 
transition seems to occur (for a sample size $1024$x$1024$) at the critical loss
$R_{c}=10^{-5}$. The correlation length $L_{c}$ at this loss is about $1024$.
Therefore, there is a scaling effect for $R_{c}$ for any finite system. We can 
then avoid this scaling effect if we consider an infinite system size. In this case, 
there is a phase transition from localized modes to extended ones at the critical loss 
$R_{c}=0$ and the conductivity distribution is log-normal only at this critical loss. 

\hspace{0.33 in} This statistical characterization of the classical eigenmodes
is exactly the inverse of that of the electronic states. Indeed, in electronic
systems, we look for the statistical behavior of the conductance while in this
case the analog of the electronic conductance is the optical transmission which
should show a similar statistical behavior. Neverthless, this characterization
of the electromagnetic eigenmodes by the statistical properties of the $ac$
conductivity is also related to that of the electronic states. Actually, we see
for vanishing losses (corresponding to extended modes) that the distribution
of the real conductivity seems to become independent of the system size (see Fig.6b)
and its relative variance is constant (Fig.5). This behavior is also shown in 
electronic systems where in the metallic regime, the conductance fluctuations 
become independent of the length and magnetic field (called universal conductance 
fluctuations \cite{Zano}).

\section{Conclusion}

\hspace{0.33 in} We have studied in this paper the characteristic lengths
and the statistical properties of the real part of the effective conductivity
in 2D metal-dielectric composites at the percolation threshold and for a
characteristic frequency where the conductivities of the two components have
the same magnitude for a vanishing loss. We found that the correlation length and the 
localization length (studied in the previous paper \cite{zekri1}) have a similar 
behavior. These lengths seem to diverge for vanishing losses which leads
to the limit of validity of the effective medium theory (since the size of the system 
is never greater than these lengths) and confirms the behavior of the real 
conductivity observed previously \cite{zekri1} in this region of the loss.

\hspace{0.33 in} We examined also the statistical properties of this effective 
conductivity and found two different distributions for small and large losses. 
For large losses
the distribution of conductivity is Gaussian while for small losses it becomes
log-normal with a long tail and strong fluctuations. For vanishing losses, the
fluctuations seem to change very slowly with the size of the system. We found the
critical loss separating these two statistical behaviors at $R_{c}=10^{-5}$. This 
critical loss separates also the localized modes at large losses from the extended 
modes at small ones \cite{zekri1}. However, this critical loss is size dependent and
actualy the correlation length at $R_{c}$ is about $1024$ which is the size of
the system. Neverthless, for infinite systems there is a phase transition from
localized modes for finite losses to extended modes at zero loss which is then
the critical loss for this phase transition. 

\hspace{0.33 in} We found an analogy between the 
statistical characterization of the classical eigenmodes studied here and that of
the electronic states. In particular, 
we found that in the limit of vanishing loss where the eigenmodes are extended,
constant  conductivity fluctuations (with the length scale) which corresponds to
the metallic states in electronic systems. Therefore, these scaling and
statistical behaviors of the real conductivity explain its anomalous decay with the
loss and its non monotonic variation observed in the previous paper
\cite{zekri1}. However, since the distributions observed here have long tails
and are not really Gaussians, an extensive study in this way is needed
by using the L\'evy distributions \cite{Levy} and examining the variation
with the loss of the exponent of the power-law tail of these distributions.
We need also an accurate determination of the correlation length in order to
study its critical divergence exponent for vanishing loss. Indeed, the divergence of 
$L_{c}$ is power-law with the exponent $0.41$ close to the estimation of Brouers 
et al. \cite{Brou3} ( $0.5$) but no accurate determination has been done because of the 
strong fluctuations of the effective conductivity. Therefore an exact calculation of 
the conductivity (by Frank and Lobb method \cite{Frank}) and its averaging by taking 
into account their statistical distributions are important for the accurate 
determination of $L_{c}$ by the above mentioned scaling behavior.  These 
questions should be a subject of a forthcoming investigation.

\vspace{0.2 in}

{\bf ACKNOWLEDGEMENTS}

\hspace{0.33 in} We would like to acknowledge the support and the hospitality of ICTP 
during the progress of this work.

\newpage

\begin{center}
{\bf Figure Captions}
\end{center}

\bigskip

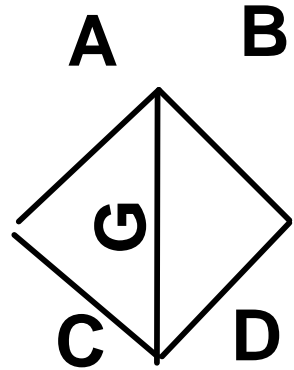
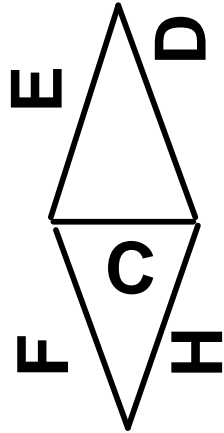
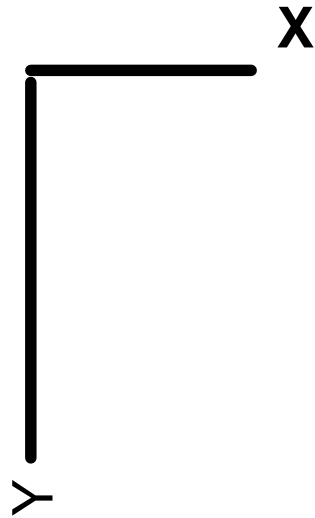
{\bf Fig.1 } The real space renormalization group for a square network.

\bigskip

{\bf Fig.2} The real part of the conductivityversus the sample size for different 
losses $R$ and for a) $\omega = 1/8$ and b) $\omega = 1$. 

\bigskip

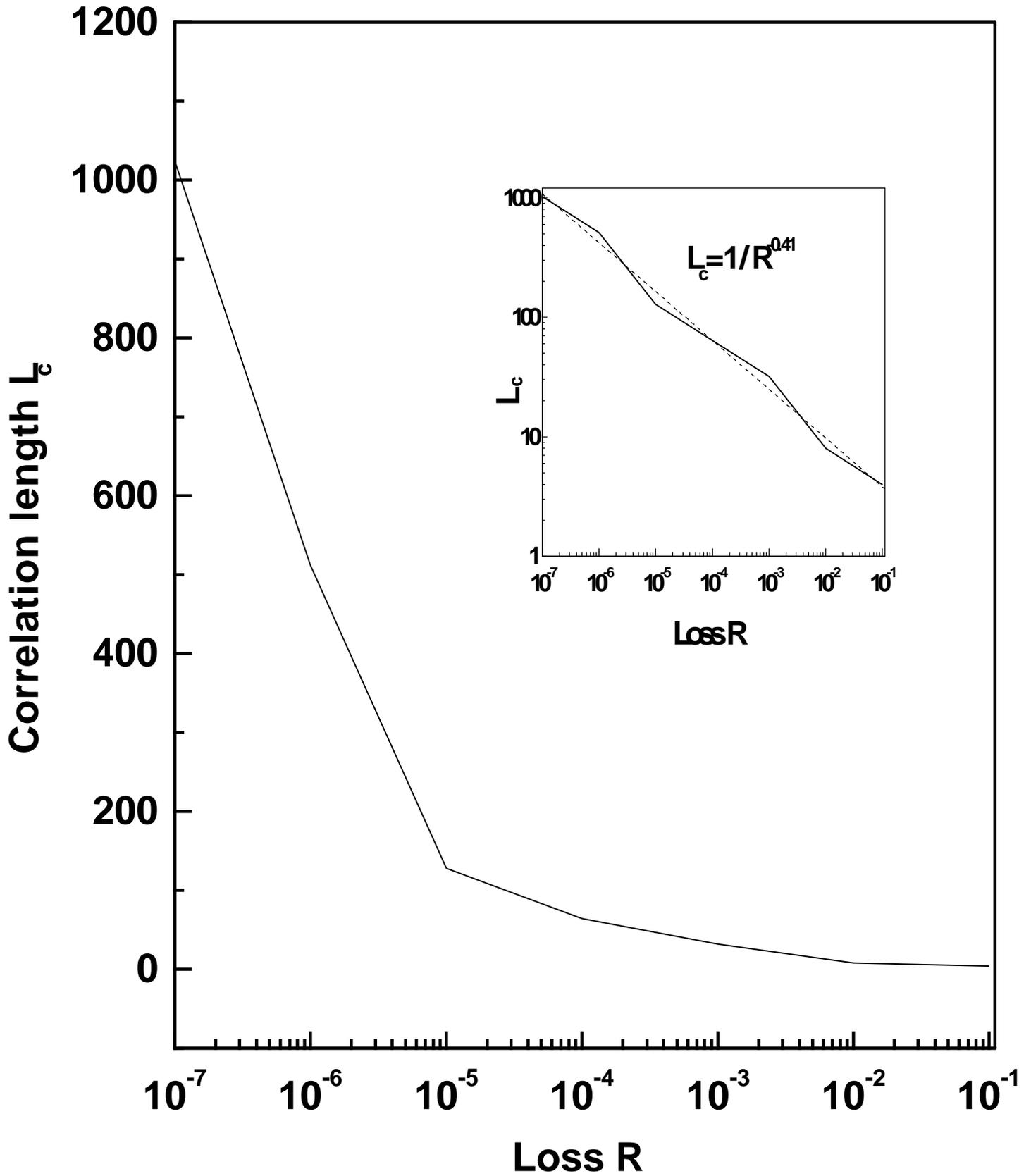
{\bf Fig.3} The correlation length as a function of the loss $R$. The insert is a 
log-log plot of this figure for the estimation of the power-law exponent.

\bigskip

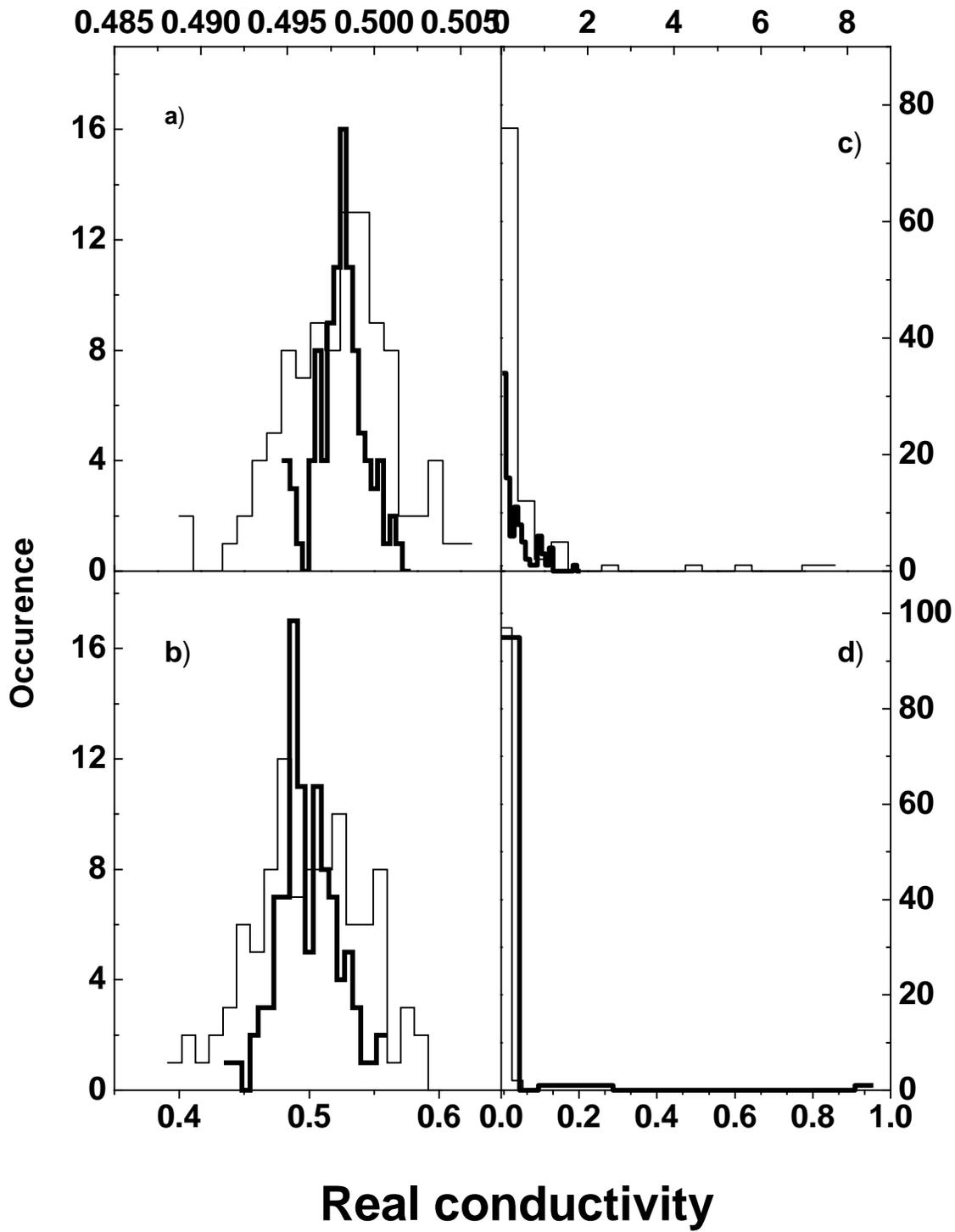
{\bf Fig.4} The distribution of the real part of the effective conductivity
for a sample size $512$x$512$ (thin curve) and $1024$x$1024$ (thick curve) and
for 4 values of the loss: a) $R=10^{-1}$, b) $R=10^{-3}$, c) $R=10^{-6}$ and
d) $R=10^{-9}$.

\bigskip

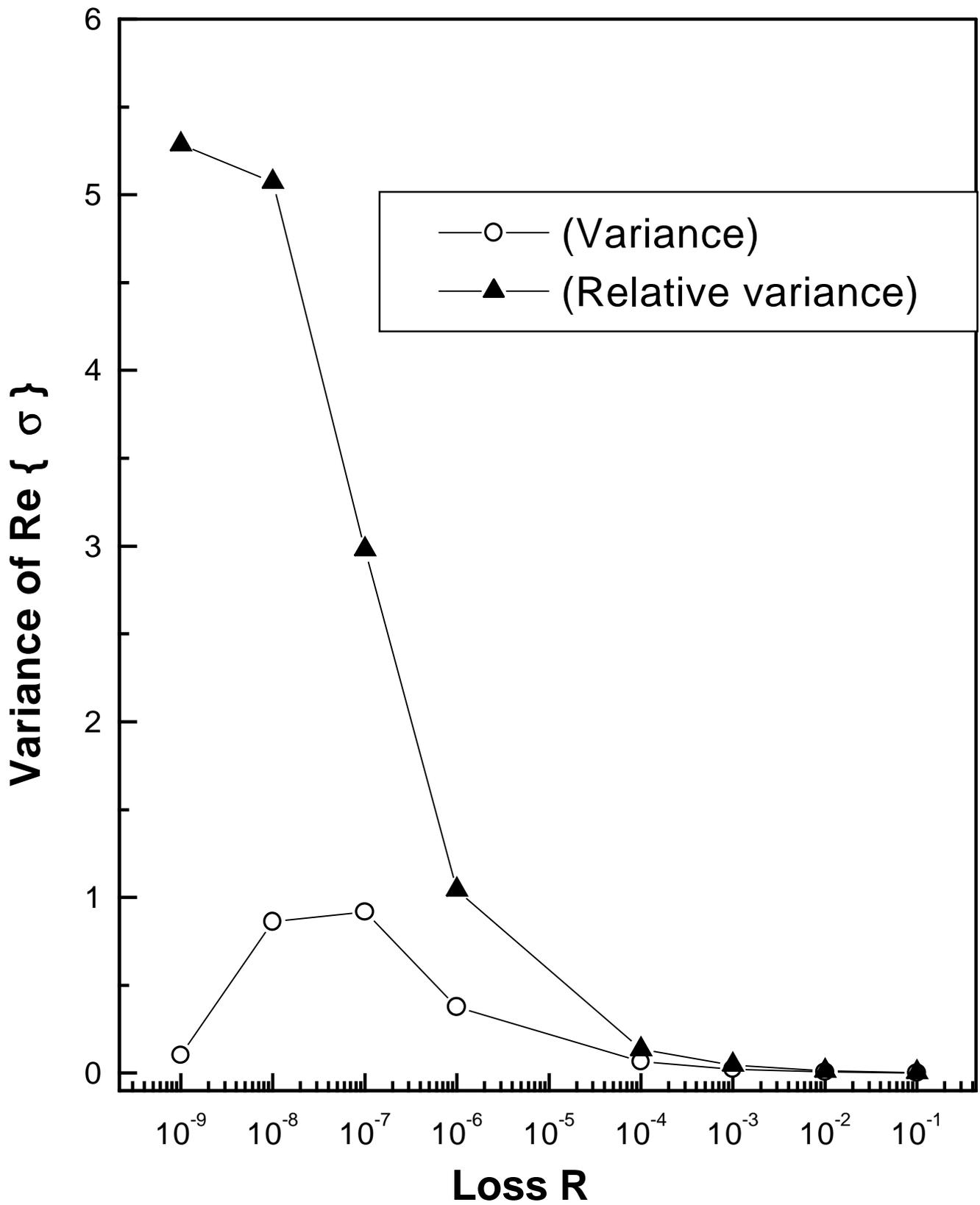
{\bf Fig.5} The variance of the real part of the effective conductivity (open circles) 
and the relation variance (filled triangle up). 

\bigskip

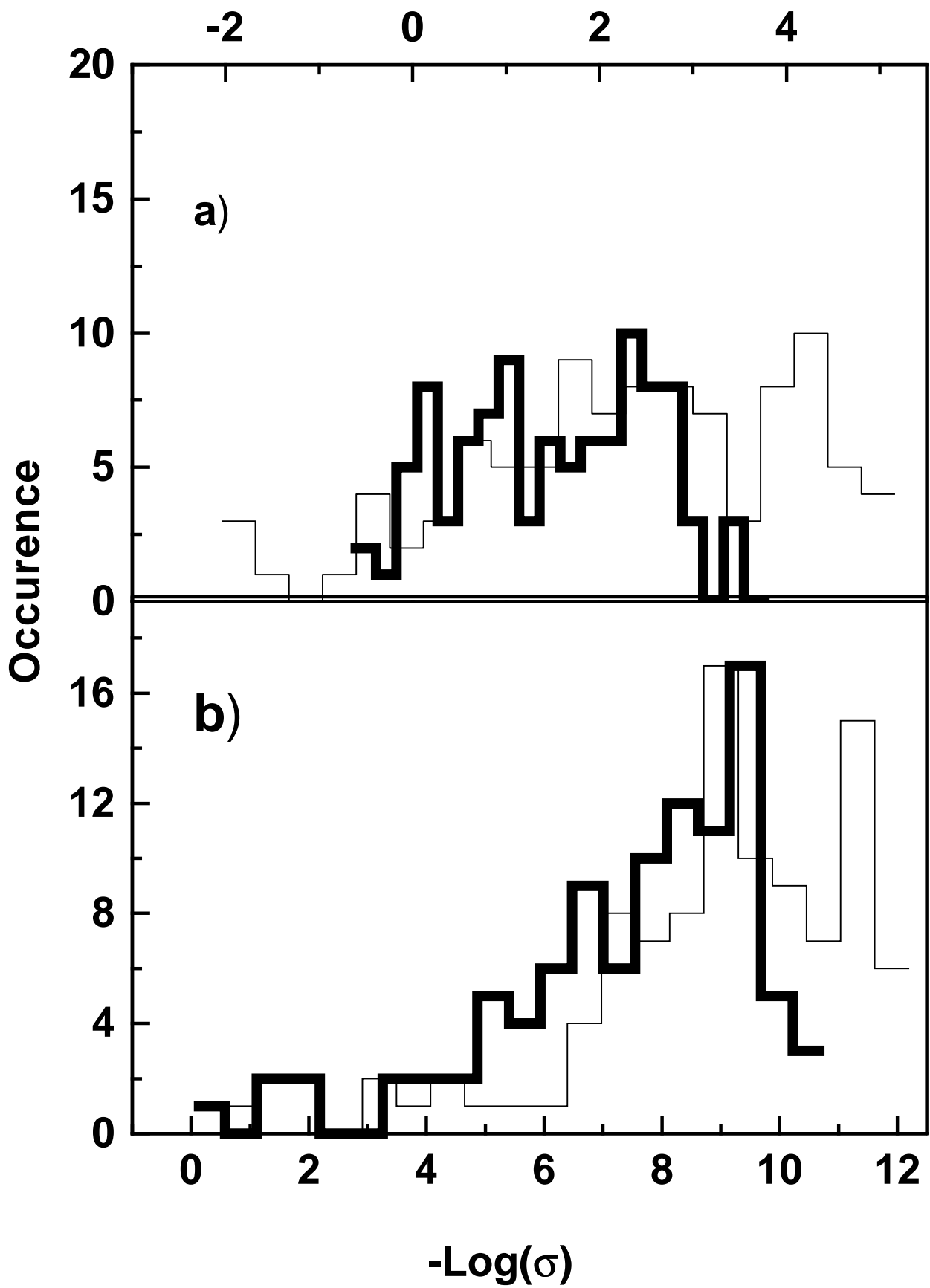
{\bf Fig.6} The distribution of $-Log( Re[ \sigma])$ for sample sizes $512$x$512$ 
(thin curve) and $1024$x$1024$ (thick curve) and for: a) $R=10^{-6}$ and b) $R=10^{-9}$.

\bigskip

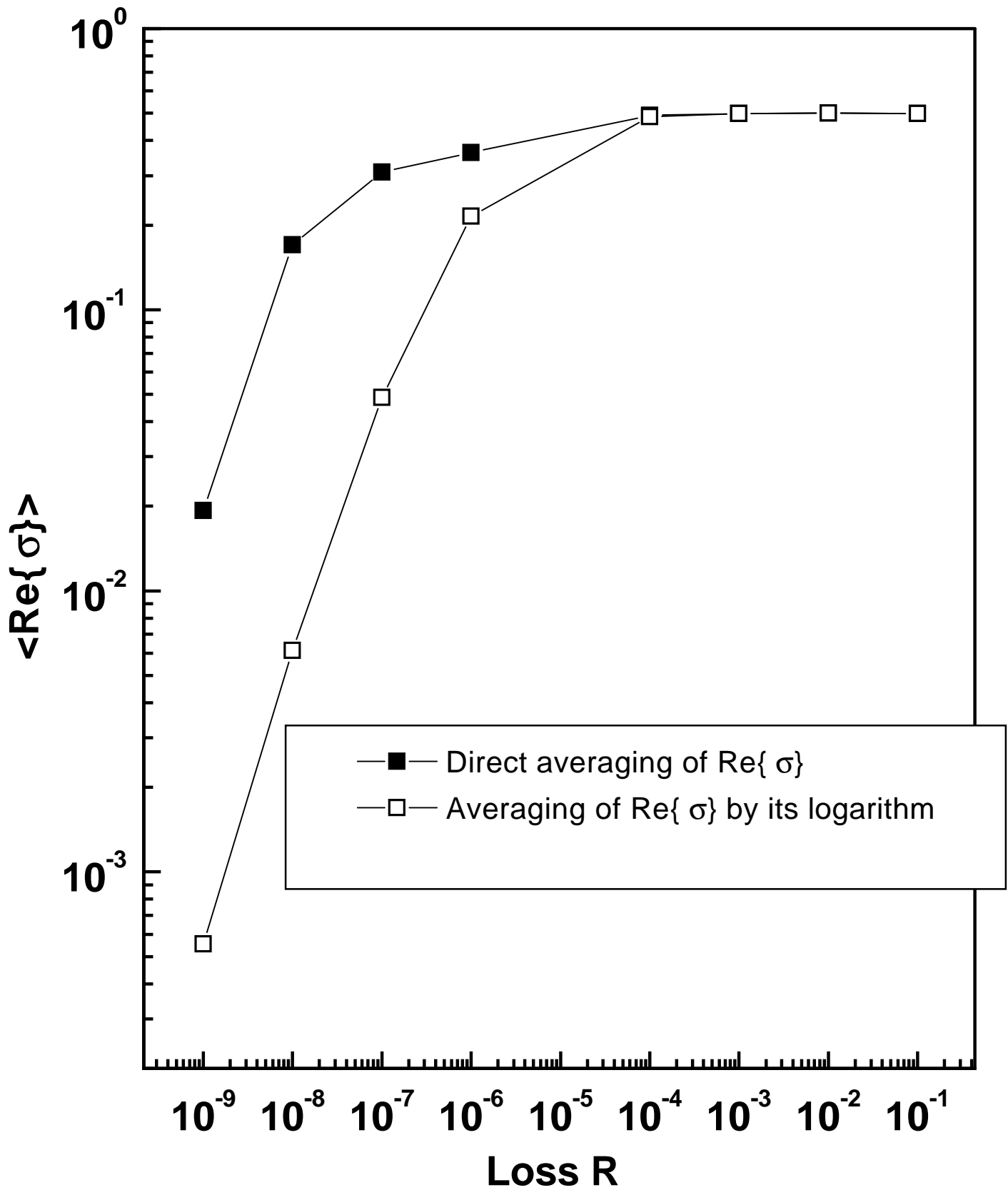
{\bf Fig.7} Averaged real part of the conductivity versus $R$ by averaging 
$Re[ \sigma ]$ (filled squares) and $Log( Re[ \sigma])$ (open squares).

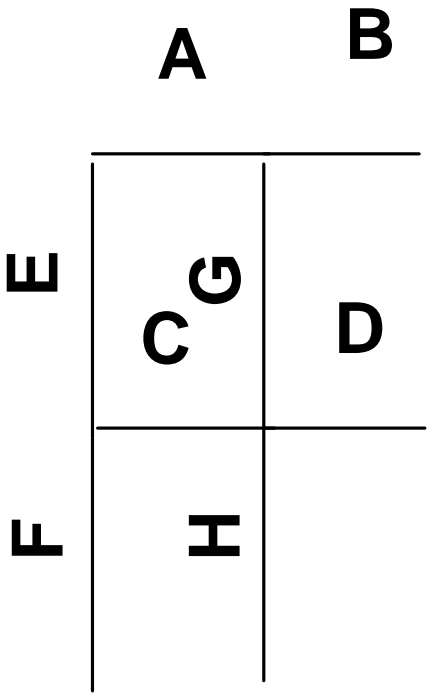
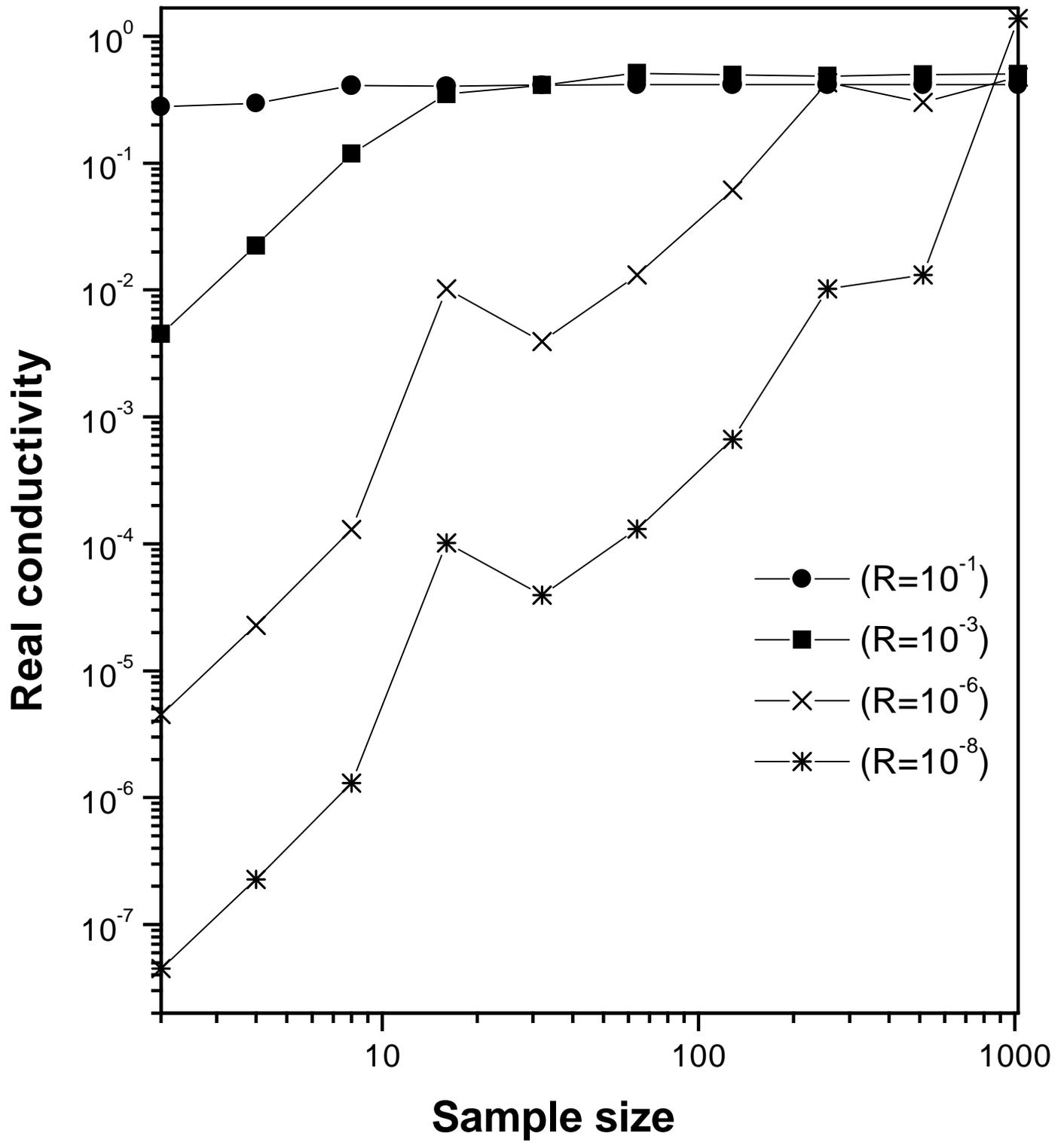
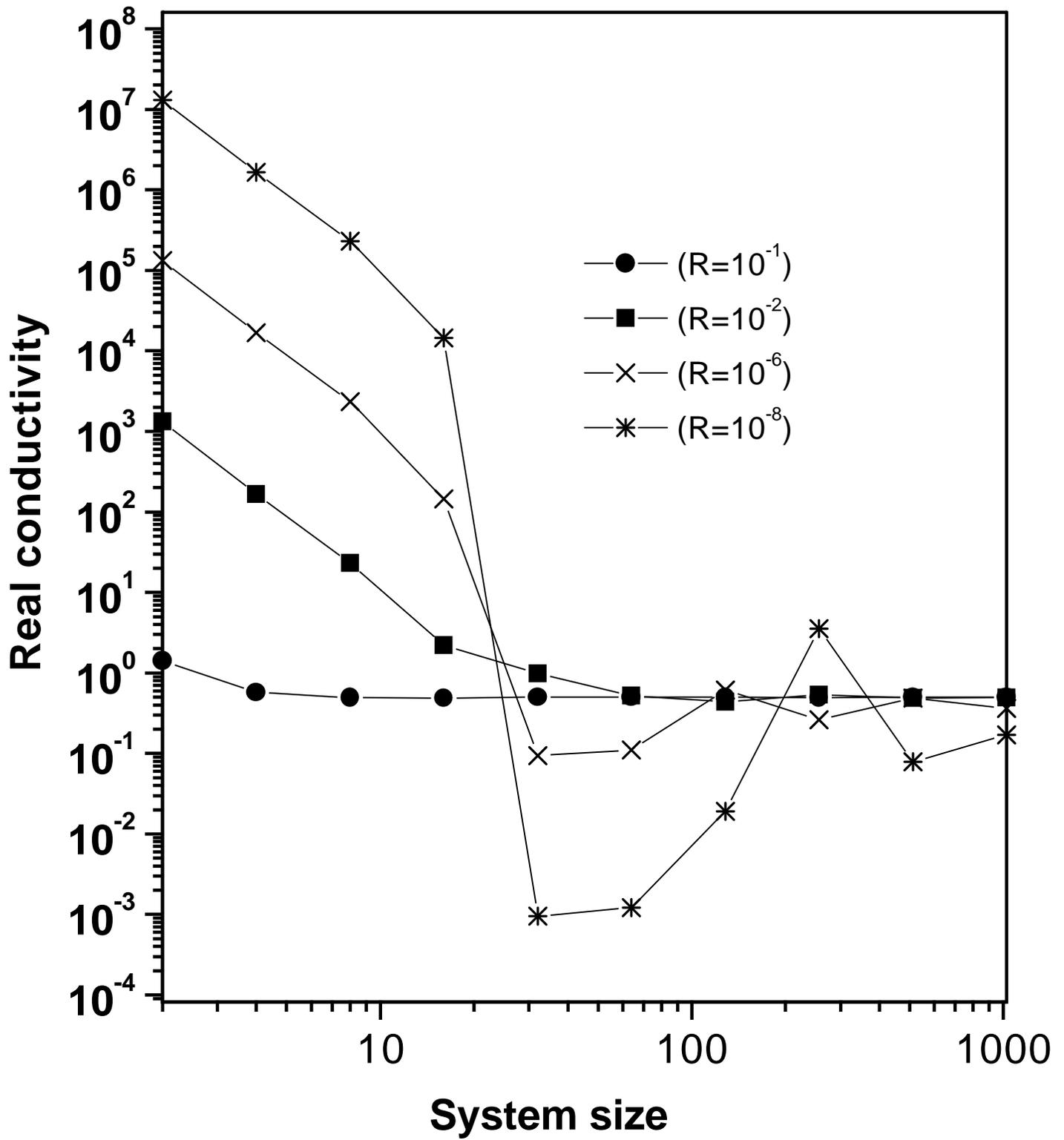
\end{document}